\documentclass[12pt]{article}

\usepackage[T1]{fontenc}
\usepackage{lmodern}

\usepackage{setspace}
\setdisplayskipstretch{1}

\usepackage[margin=1.25in]{geometry}

\usepackage[dvipsnames]{xcolor}
\usepackage{pdfpages}
\usepackage{float}
\usepackage{multirow}
\usepackage{caption}
\usepackage{subcaption}
\usepackage{epigraph}
\usepackage{titlesec}

\titleformat{\section}
  {\normalfont\large\bfseries}{\thesection}{1em}{}
\titleformat{\subsection}
  {\normalfont\normalsize\bfseries}{\thesubsection}{1em}{}

\usepackage{mathtools}
\usepackage{amssymb}
\usepackage{amsthm}
\usepackage{amsfonts}
\usepackage{aliascnt}
\usepackage{accents}
\usepackage{dutchcal}
\usepackage{bbm}

\usepackage{enumitem}
\usepackage{sgame}
\usepackage{tikz}
\usepackage{tikz-cd}
\usetikzlibrary{calc,shapes,arrows}
\usepackage{tcolorbox}
\usepackage{listings}
\usepackage[normalem]{ulem}

\usepackage[round]{natbib}
\newcommand{\R}{\mathbf{R}}

\newcommand{\Prb}{\mathbb{P}}

\newcommand{\CE}{\mathsf{CE}}

\newcommand{\K}{\mathcal K}
\newcommand{\V}{\mathcal V}

\DeclareMathOperator{\Eq}{\mathsf{Eq}}

\setlist[enumerate]{label=\textnormal{(\roman*)},itemsep=3pt,topsep=5pt}

\newtheorem{theorem}{Theorem}[section]

\newaliascnt{proposition}{theorem}
\newtheorem{proposition}[proposition]{Proposition}
\aliascntresetthe{proposition}

\newaliascnt{lemma}{theorem}
\newtheorem{lemma}[lemma]{Lemma}
\aliascntresetthe{lemma}

\newaliascnt{corollary}{theorem}
\newtheorem{corollary}[corollary]{Corollary}
\aliascntresetthe{corollary}

\newaliascnt{claim}{theorem}

\aliascntresetthe{claim}

\theoremstyle{definition}

\newaliascnt{definition}{theorem}

\aliascntresetthe{definition}

\newaliascnt{example}{theorem}

\aliascntresetthe{example}

\newaliascnt{assumption}{theorem}

\aliascntresetthe{assumption}

\newtheorem{axiom}{Axiom}

\newaliascnt{condition}{theorem}

\aliascntresetthe{condition}

\newaliascnt{question}{theorem}

\aliascntresetthe{question}

\newaliascnt{remark}{theorem}

\aliascntresetthe{remark}

\newaliascnt{remarks}{theorem}

\aliascntresetthe{remarks}

\newaliascnt{aside}{theorem}

\aliascntresetthe{aside}

\newaliascnt{note}{theorem}

\aliascntresetthe{note}

\usepackage{thmtools}
\usepackage{thm-restate}

\usepackage{hyperref}

\hypersetup{
    colorlinks=true,
    linkcolor=OrangeRed,
    filecolor=Thistle,
    urlcolor=Thistle,
    citecolor=Thistle,
}

\usepackage[nameinlink]{cleveref}

\crefname{theorem}{theorem}{theorems}
\Crefname{theorem}{Theorem}{Theorems}
\crefname{proposition}{proposition}{propositions}
\Crefname{proposition}{Proposition}{Propositions}
\crefname{lemma}{lemma}{lemmas}
\Crefname{lemma}{Lemma}{Lemmas}
\crefname{corollary}{corollary}{corollaries}
\Crefname{corollary}{Corollary}{Corollaries}
\crefname{claim}{claim}{claims}
\Crefname{claim}{Claim}{Claims}
\crefname{definition}{definition}{definitions}
\Crefname{definition}{Definition}{Definitions}
\crefname{example}{example}{examples}
\Crefname{example}{Example}{Examples}
\crefname{assumption}{assumption}{assumptions}
\Crefname{assumption}{Assumption}{Assumptions}
\crefname{axiom}{axiom}{axioms}
\Crefname{axiom}{Axiom}{Axioms}

\makeatletter
\let\cref@old@isrefconsecutive\cref@isrefconsecutive
\def\cref@isrefconsecutive#1#2{%
  \begingroup
    \def\cref@assumptiontype{assumption}%
    \cref@gettype{#1}{\cref@typea}%
    \ifx\cref@typea\cref@assumptiontype
      \endgroup
      \@cref@refconsecutivefalse
    \else
      \endgroup
      \cref@old@isrefconsecutive{#1}{#2}%
    \fi
}
\makeatother

\crefname{condition}{condition}{conditions}
\Crefname{condition}{Condition}{Conditions}
\crefname{question}{question}{questions}
\Crefname{question}{Question}{Questions}
\crefname{remark}{remark}{remarks}
\Crefname{remark}{Remark}{Remarks}
\crefname{remarks}{remarks}{remarks}
\Crefname{remarks}{Remarks}{Remarks}
\crefname{aside}{aside}{asides}
\Crefname{aside}{Aside}{Asides}
\crefname{note}{note}{notes}
\Crefname{note}{Note}{Notes}
\crefname{appendix}{appendix}{appendices}
\Crefname{appendix}{Appendix}{Appendices}

\newcommand{\secref}[1]{\hyperref[#1]{\S\ref*{#1}}}
\newcommand{\conditionref}[1]{\hyperref[#1]{\textcolor{Blue}{\textnormal{\ref*{#1}}}}}
\newcommand{\Conditionref}[1]{\hyperref[#1]{\textcolor{Blue}{\textnormal{\ref*{#1}}}}}

\definecolor{backcolour}{rgb}{0.63,0.79,0.95}
\lstdefinestyle{mystyle}{
  backgroundcolor=\color{backcolour},
  basicstyle=\ttfamily\footnotesize,
  breakatwhitespace=false,
  breaklines=true,
  captionpos=b,
  keepspaces=true,
  numbers=left,
  numbersep=5pt,
  showspaces=false,
  showstringspaces=false,
  showtabs=false,
  tabsize=2
}
\begin{document}
\title{What it is to Obey}
\author{Mark Whitmeyer\thanks{For KS.}}
\date{\today}
\maketitle
\begin{abstract}
Prior to agents taking part in a simultaneous-move game, a designer commits a device that sends signals to each of the agents, evaluating the resulting set of equilibria according to a (Hausdorff-)continuous function. In games of incomplete information I show that it must be without loss of optimality for the designer to restrict attention to direct obedient devices if and only if she assumes her preferred equilibrium will be played. In games of complete information, I characterize the monotone evaluations, which include both the maximally optimistic one and the maximally pessimistic one. 
\end{abstract}

Simply ``tell them what to do in such a way that they would be willing to obey'' says the obedience principle, a fundamental feature of mechanism and information design. Indeed, the concepts of correlated equilibrium and Bayes correlated equilibrium largely take this principle as given.

But take the perspective of a principal, who designs the correlating (and, possibly, information-providing) device. She has preferences over the profile of actions played in the game, evaluating her set of payoffs generated by equilibria in a game according to a function \(\V\) whose domain is the set of nonempty compact sets of real numbers \(\mathcal{K}(\R)\) (and which I also assume is (Hausdorff-) continuous and \textit{internal}\footnote{Meaning that \(\V\) produces an output lying between the worst and best equilibrium payoffs}). If the obedience principle holds in all (finite) games of incomplete information--if it is without loss of optimality for the designer to restrict attention to obedient direct devices--what does that say about \(\V\)?

That is, suppose it is without loss of optimality for the principal to restrict attention to direct obedient devices in every environment (tuple of finite states, common full-support prior, utilities and finite sets of actions for finitely-many agents, and principal's state dependent utility functions). Then what form must any continuous, internal \(\V\) take? 

I argue that \textbf{it must be the case that \(\V = \max\)}. And as the obedience principle classically holds under design-optimal equilibrium selection, this is a complete characterization of \(\V\)s for which the obedience principle holds.

To prove necessity, I construct a family of two-player coordination games in which the unknown state is two dimensional, with one corresponding to an agent's ``target" and the other his ``role.'' In these games, merely recommending actions necessarily leaves the agents uncertain about their roles, allowing for additional equilibria that yield the principal lower payoffs than in her most-preferred equilibrium. An indirect device that also reveals to the agents their roles makes the desired behavior the unique equilibrium. Accordingly, this device is strictly better for the principal for any evaluator other than \(\max\). In a sentence, the loss of information by moving from an indirect to a direct device can always be made harmful, as it opens up opportunities to coordinate in damaging ways.

That was for Bayes-correlated equilibrium, apposite to games of incomplete information. What about games of complete information, where the only role of the device is to correlate? Clearly, \(\V = \max\) is now sufficient for the obedience principle to hold. \textbf{But now so too is \(\V = \min\)!} Indeed, observe that any set of equilibria under a device must include the Nash equilibria (the device's output is just cheap talk after all), and those can always be made obedient for some direct device.

In fact, the set of obedience-principle-satisfying evaluations is larger, even when I restrict attention to monotone ones, which require the evaluation to increase whenever the principal's utility of actions increases pointwise. I show that for such monotone (and continuous and internal) \(\V\)s, the obedience principle holds if and only if \(\V(S)=\max\{\min S,h(\max S)\}\) for every \(S\in\K(\R)\), for some increasing and continuous $h\colon\R\to[-\infty,\infty)$ satisfying $h(b)\leq b$ for every $b$ and $h(b)=\inf_{a\leq b}\V(\{a,b\})$.

The necessity argument has two parts. I first show that monotonicity of the evaluation forces it to depend only on the best and worst equilibrium payoffs. I then (similar to the incomplete information scenario) construct games in which an indirect device avoids a low-payoff equilibrium that every direct device approaching the best payoff must engender. These games imply that, whenever the principal's evaluation exceeds her worst payoff, lowering that payoff while holding the best one fixed cannot change her evaluation, as otherwise she'd strictly prefer an indirect device to every direct one. In sum, the principal can care about a deterioration of her worst equilibrium only when she evaluates the device by that worst equilibrium.

For sufficiency, as I noted above, the principal cannot escape the Nash-equilibrium payoffs. On the other hand, she obviously cannot do better than her favorite correlated-equilibrium payoff. Consequently, the larger of the worst Nash-equilibrium payoff and \(h\) of the highest correlated-equilibrium payoff bounds the principal's evaluation. If the worst Nash-equilibrium payoff determines this bound, well, she can always attain it via an obedient direct device. Otherwise, recommendations from a correlated equilibrium yielding her the highest payoff attain the bound. In either case, an obedient direct device is optimal.

\subsection*{Related Work}

The standard obedience principle allows a principal to reproduce an equilibrium outcome of an arbitrary information device as the obedient equilibrium outcome of a direct device.\footnote{See \citet{Aumann1974} and \citet{Aumann1987} for correlated equilibrium, and \citet{BergemannMorris2016} for Bayes correlated equilibrium.} The direct device need not reproduce the original device's other equilibria, nor does the principle (or principal) rule out additional equilibria engendered by the direct device.\footnote{Direct recommendations can also fail to preserve equilibrium refinements. \citet{DhillonMertens1996} establish this for perfect correlated equilibrium, while \citet{LuoQiaoSun2022} obtain a revelation principle for an alternative refinement allowing message-dependent trembles.} I ask when a principal who evaluates the full set of equilibrium payoffs can nevertheless restrict attention to obedient recommendations.

\citet{MurraAntonRubbiniSerrano2025} also characterize when a recommendation principle holds. They allow a receiver's utility to depend nonlinearly on her beliefs and identify conditions under which pooling messages that induce the same action preserves its optimality. Their requirement (in contrast to mine) is the traditional one--that \textit{some} optimal response induces the desired outcome.

Closer yet are the papers that prescribe or study an information designer's treatment of equilibrium multiplicity. Some impose selection of her favorite equilibrium; others, adversarial selection of her least favorite one.\footnote{\citet{Taneva2019} studies designer-optimal information design in games. \citet{BM2019} explain why adversarial equilibrium selection can invalidate the usual obedience argument. \citet{MPT2020} allow for different selection rules and study an investment game under adversarial selection; see also \citet{MorrisOyamaTakahashi2024}, who study information design under adversarial selection in binary-action supermodular games.} \citet{BMV2024} endow their designer with an arbitrary Hausdorff-continuous objective over the set of equilibrium outcomes and show that simple information structures suffice for a broad class of information-design problems. \citet{Zeng2026} provides conditions under which a designer can preserve a perfect Bayesian equilibrium outcome by recommending actions together with information about the agents' conjectures, when their responses must respect specified selection restrictions.

Earlier work also studies the additional equilibria of correlation devices. Of special note is \citet{Ray2002} who constructs a two-player game in which both obedience and disobedience are equilibria of a direct device. He constructs a non-canonical device that induces the obedient outcome but under which the corresponding disobedient response profile is not an equilibrium.\footnote{See also \citet{Ray2008}.} My complete-information construction builds off of his two-player game and a device from his family. I establish further that no equilibrium of this device induces the unwanted action distribution, then add a third agent whose best responses under this device deliver the principal prescribed payoffs at the first two agents' equilibrium action distributions, without changing their incentives.\footnote{I use the resulting three-agent game to show that every direct device that comes close to the principal's highest attainable equilibrium payoff also has an equilibrium giving her less than the indirect device's worst equilibrium payoff.}

Finally, full implementation in mechanism design requires the full set of equilibrium outcomes to coincide with a desired set.\footnote{\citet{PalfreySrivastava1989} proffer an example of an allocation rule that an indirect mechanism fully implements in spite of every equivalent direct mechanism having additional equilibrium outcomes. \citet{KarRaySerrano2010} show that selection rules for a class of correlated equilibria violate Maskin monotonicity and cannot be fully implemented in Nash equilibrium.} In a companion paper to this one, \citet{revprinc2026}, I characterize when the revelation principle holds universally. Without transfers, like here, the principal's evaluation must be \(\max\); with, it is (approximately) irrelevant.

\section{Intuition}

I start with the easier case; games of incomplete information. For them, the key argument is contained in \Cref{lem:bce-construction}. For simplicity, consider a binary set of payoffs \(F=\{b,a\}\) (\(a<b\)). I construct a game of incomplete information with two agents in which every obedient direct device yielding the principal a payoff of \(b\) also induces a strict equilibrium that produces a payoff \(a\) for the principal, yet there exists a non-direct device that yields the principal a payoff of \(b\) in the unique equilibrium.

The state has two dimensions. One assigns one of two available target actions to each of the agents; the other designates the role of each agent. The first role gives an agent payoff \(0\) from choosing his target and \(-1\) from choosing the other action, regardless of the other agent's choice:
\[
\begin{array}{c|cc}
 & \text{target} & \text{other action}\\ \hline
\text{target}
 & (0,0\,;\,b)
 & \left(0,-1\,;\,\dfrac{3a-b}{2}\right)\\[2mm]
\text{other action}
 & \left(-1,-2\,;\,\dfrac{3a-b}{2}\right)
 & (-1,1\,;\,a).
\end{array}
\]
And symmetrically if the other agent has this role.

After averaging the agents' utilities (over the two equally likely roles):
\[
\begin{array}{c|cc}
 & \text{target} & \text{other action}\\ \hline
\text{target}
 & (0,0\,;\,b)
 & \left(-1,-1\,;\,\dfrac{3a-b}{2}\right)\\[2mm]
\text{other action}
 & \left(-1,-1\,;\,\dfrac{3a-b}{2}\right)
 & (0,0\,;\,a).
\end{array}
\]
Any direct device \(q^0\) that yields the principal her maximal payoff of \(b\) must instruct each agent to choose his target. Such a device, however, cannot reveal their roles (as it would no longer then be direct). Moreover, every other equilibrium in response to the device gives the principal a payoff of \(a\).

The second device, \(\lambda^+\), also reveals the agents' roles. Once they are revealed, the agent with the first
role strictly prefers his target regardless of the other agent's
choice. Given that choice, the other agent also strictly prefers his target. Hence, \(\lambda^+\) results in a unique equilibrium, in
which both agents choose their targets and the principal receives
\(b\).

This construction is easy to extend to an arbitrary finite set of payoffs \(F=\{b=a_0>a_1>\cdots>a_k\}\).\footnote{With the extension to compact sets following from the strictness of the equilibria I construct, compactness of the various spaces, and continuity of the evaluator \(\V\).} Now each agent has \(k+1\) targets and actions, arranged around a circle. I construct the agents' utilities so that when the direct device reveals the targets but not the roles (as they would in any \(b\)-attaining direct device), they produce incentives to coordinate: each agent wants to minimize the expected squared distance between his difference (between his action and his target calculated modulo \(k+1\)) and the other agent's difference. This yields \(k+1\) pure strategy equilibria in which each chooses the same difference, with each agent’s unconditional distribution of differences being uniform over two adjacent integers in every other equilibrium. I choose the principal's payoffs so that these pure-strategy equilibria exhaust the payoffs in \(F\) (and are strict) and the remaining ones produce the payoffs in \(F\) other than \(b\). Revealing the roles again makes choosing the target the unique equilibrium and delivers the principal \(b\).

With this lemma in hand, \Cref{thm:bce-characterization} is almost immediate. Sufficiency is classical. Toward necessity, suppose that \(\V \neq \max\). Then there is then a nonempty compact set \(S\) such that \(\V(S)<\max S\). By continuity of \(\V\), I can choose a finite set \(F\), with maximum \(b\), such that \(\V(F)<b\). \Cref{lem:bce-construction} supplies a device \(\lambda^+\) with a unique equilibrium giving the principal \(b\), whereas every direct device has value bounded strictly below \(b\).

\subsection{Complete information}

For games of complete information, the key argument is contained in two
lemmas, \Cref{lem:ce-endpoints,ce:construction}. In the first, I use monotonicity of the evaluation \(\V\) to simplify the form
it can take and thereby single out binary-payoff sets as the ``test
sets'' needed to characterize \(\V\). To elaborate, I show that \(\V\)
satisfies monotonicity if and only if there is a continuous
function $f$ on
$\{(m,M)\in\R^2\colon m\leq M\}$, increasing in each argument and
satisfying $m\leq f(m,M)\leq M$, such that
\(\V(S)=f(\min S,\max S)\) for every \(S\in\K(\R)\). In that case,
$f(m,M)=\V(\{m,M\})$.

In the proof, I take the following two-player game (depicted in normal form)--which has exactly \(k\) equilibria for whatever \(k\) I want--and the trivial device which outputs constant signals.
\[
\begin{array}{c|ccccc}
 & 1 & 2 & \cdots & k-1 & k\\ \hline
1
  & (1,1) & (0,0) & \cdots & (0,0) & (0,1)\\
2
  & (0,0) & (1,1) & \cdots & (0,0) & (0,1)\\
\vdots
  & \vdots & \vdots & \ddots & \vdots & \vdots\\
k-1
  & (0,0) & (0,0) & \cdots & (1,1) & (0,1)\\
k
  & (1,0) & (1,0) & \cdots & (1,0) & (1,1).
\end{array}
\]
With this game in hand, I vary only the principal's payoffs. Under one, the equilibria \((1,1),\ldots,(k,k)\) give her \((x_1,\ldots,x_k)\), where
\(m=x_1\leq\cdots\leq x_k=M\). Under another, she gets \((m,\ldots,m,M)\); and under a third, \((m,M,\ldots,M)\); while all three give her \(0\) off the diagonal. Since \((m,\ldots,m,M)
 \leq (x_1,\ldots,x_k)
 \leq (m,M,\ldots,M)\), monotonicity means the principal's evaluation of
\(\{x_1,\ldots,x_k\}\) bust nest between her evaluations of the other two. Moreover, both comparators generate the identical set of payoffs \(\{m,M\}\), forcing \(\V(\{x_1,\ldots,x_k\})=\V(\{m,M\})\). I then use continuity of \(\V\) to extend this conclusion from finite to compact sets. The converse is nearly immediate: if \(\V\) depends only on the minimum and maximum
through a function \(f\) that increases in each argument, then raising the principal's payoff at every action profile raises both
endpoints of her equilibrium payoff set and cannot lower her
evaluation.

The other lemma I need, \Cref{ce:construction}, is more involved. There, I construct a game with several useful features. Start with a binary set of principal payoffs $\{a,b\}$, with \(c < a < b\). There are two agents, 1 and 2, who play a binary anti-coordination game with three Nash equilibria: two strict pure ones, in which they miscoordinate; and a mixed-strategy one. I then choose a direct device $q^*$ under which both obedience and (always) being disobedient are strict equilibria, the latter inducing the distribution over actions $q^-$. I also construct an indirect device $\lambda_0$ that has an equilibrium that induces $q^*$ but no equilibrium that induces $q^-$,\footnote{The new device can be decomposed into two parts. First, one of two public components is revealed. Second, in one of the two realized components, agents receive nothing more so the agents play one of the Nash equilibria of the game. In the other, agents receive additional private recommendations, chosen so that the possible best responses to them, plus the Nash-equilibrium behavior in the other component, place an upper bound on the probability with which a particular action profile can be played, strictly less than that under $q^-$.} and whose remaining equilibria induce only finitely many other action distributions.

I then construct a function $F$ of the two agents' action distribution such that $F(q^-)=c$, $F(q^*)=b$, and $F=a$ at every other equilibrium distribution generated by $\lambda_0$ (including at every Nash equilibrium distribution of the game).\footnote{Because $\lambda_0$ generates only finitely many equilibrium distributions and none equals $q^-$, I can choose disjoint neighborhoods of $q^*$ and $q^-$ containing no other such distributions, let $F$ equal $a$ outside them, and let it rise to $b$ at $q^*$ and fall to $c$ at $q^-$. More specifically, I choose $F$ so that it is the difference of two maxima of finitely many affine functions.} To make these the principal's actual payoffs, I then modify the game by adding a third agent: whenever the first two agents' actions have distribution $P$, every best response of the third agent yields the principal payoff $F(P)$. Agents 1 and 2 retain their original payoffs, which do not depend on agent 3's action. I extend $\lambda_0$ to the three-agent game by giving agent 3 a constant signal, which now produces payoff set for the principal $\{a,b\}$, while every Nash equilibrium delivers the principal payoff $a$.

The key distinction is that a direct device cannot generate the high payoff $b$ without also producing the low payoff $c$. If an equilibrium of a direct device gives the principal payoff $b$,
then no matter what agent 3 observes, the first two agents'
action distribution must be $q^*$. Otherwise, since $q^*$ uniquely
maximizes $F$, the principal's average payoff would be strictly below $b$. But doing the opposite then induces $q^-$. Since the first two agents' incentives hold strictly in this equilibrium, the same conclusion holds approximately for every direct device whose highest
equilibrium payoff is sufficiently close to $b$: such a device must
also have an equilibrium whose payoff is close to $c$.

With these lemmas in hand, I prove the theorem. For necessity, \Cref{lem:ce-endpoints} reduces the problem to
understanding how \(f(m,M)\) changes with the worst payoff \(m\) when
the best payoff \(M\) remains fixed. I first take \(c<a<B\) and show that, if \(f(a,B)>a\), lowering the
worst payoff from \(a\) to \(c\), while keeping the best payoff \(B\)
fixed, cannot change the evaluation. Suppose instead that
\(f(c,B)<f(a,B) \eqqcolon r\) and let \(b\) be the smallest
element of \([a,B]\) such that \(f(a,b)=r\). \Cref{ce:construction} then supplies an environment containing a device whose payoff set is \(\{a,b\}\). This yields the principal value \(r\), which is also optimal among all devices, as every device in this environment has \(a\) among its equilibrium payoffs and has no equilibrium payoff outside \([c,b]\).

But no direct device can attain this value, not even approximately. If its highest equilibrium payoff is close to \(b\),
\Cref{ce:construction} forces another equilibrium payoff close to
\(c\), in which case monotonicity and continuity of \(f\) plus \(f(c,B)<r\) imply its value is then strictly below \(r\). If its highest
equilibrium payoff is bounded away from \(b\), the minimality of \(b\)
again places its value strictly below \(r\). Accordingly, for each \(B\), \(f(m,B)\) either equals \(m\) or, once it exceeds \(m\), remains unchanged as \(m\) falls: in notation, defining \(h(B)\coloneqq \inf_{m\leq B}f(m,B)\), \(f(m,B)=\max\{m,h(B)\}\), with the monotonicity, continuity, and internality of \(f\) yielding the other properties of \(h\).

For sufficiency, take some environment and let \(a\) be the principal's worst
Nash-equilibrium payoff and \(B\) be her highest
correlated-equilibrium payoff. No matter the device, the agents can
ignore their signals and play any Nash equilibrium, so the principal's worst payoff is no higher than \(a\). Her best
equilibrium payoff under any device is clearly no higher than \(B\). Accordingly, \(\max\{a,h(B)\}\) is an upper bound on the principal's value for any device. An obedient direct device attains it. If \(h(B)\leq a\), simply use the direct device generated by the Nash equilibrium play that yields the principal \(a\). If \(h(B)>a\), use an obedient direct device whose obedient equilibrium delivers the principal payoff \(B\).

\section{The Design Problem}\label{sec:model}

There is a principal (she) and a finite set of agents,
$\{1,\ldots,n\}$, ($n\geq1$). Each agent $i$ simultaneously chooses
an action from a finite nonempty set $A_i$. Write
$A=\times_iA_i$. There is a state $\theta$ that belongs to a finite nonempty set $\Theta$ and is drawn from a full-support common prior $\mu\in\Delta^{\circ}(\Theta)$. If action profile \(a\) is played in state \(\theta\), agent $i$ obtains utility
$u_i(\theta,a)$ and the principal gets payoff $v(\theta,a)$.
The tuple $E=(\Theta,\mu,(A_i,u_i)_i,v)$ is the \textit{environment}, which is common knowledge.

The principal chooses finite sets of signals \(T_i\) and commits to a \textit{device} (experiment) $\lambda(t\mid\theta)$ on $T=\times_iT_i$ before nature draws the state, which privately sends each agent his signal. The agents then choose actions simultaneously. A \textit{response strategy} for player \(i\) is $\sigma_i\colon T_i\to\Delta(A_i)$. A \textit{response equilibrium} is a Bayesian Nash equilibrium in response strategies.\footnote{A response equilibrium is \textit{pure} if each $\sigma_i(t_i)$ is degenerate. A pure response equilibrium is \textit{strict} if, for every agent and every signal that occurs with strictly positive probability, the agent strictly prefers not to deviate.} I write $\Eq(\lambda)$ for the set of response equilibria. Writing $\sigma(t)=\times_i\sigma_i(t_i)$, at equilibrium, no agent has a profitable deviation:
\begin{equation}\label{eq:bce-response}
 \sum_{\theta,t_{-i}}\mu(\theta)\lambda(t_i,t_{-i}\mid\theta)
 \left[u_i(\theta,\sigma(t))
       -u_i(\theta,b_i,\sigma_{-i}(t_{-i}))\right]\geq0
\end{equation}
for every $i,t_i$, and $b_i\in A_i$. The principal's payoff
and set of payoffs are
\[
 \Pi_v(\lambda,\sigma)
 =\sum_{\theta,t}\mu(\theta)\lambda(t\mid\theta)
      v(\theta,\sigma(t))
 \qquad \text{\&} \qquad
 S_v(\lambda)=\{\Pi_v(\lambda,\sigma)\colon \sigma\in\Eq(\lambda)\},
\]
omitting the subscript $v$ when the principal's payoff is fixed. $S(\lambda)$ is nonempty and compact.

A \textit{direct device }has $T_i=A_i$ and specifies 
$q(a\mid\theta)$. It is \textit{obedient} if following the recommendations
is an equilibrium. Equivalently,
\begin{equation}\label{eq:bce-obedience}
 \sum_{\theta,a_{-i}}\mu(\theta)q(a_i,a_{-i}\mid\theta)
 [u_i(\theta,a_i,a_{-i})-u_i(\theta,b_i,a_{-i})]\geq0
\end{equation}
for every $i,a_i,b_i$. If this holds, the joint distribution over states and actions $\psi(\theta,a)=\mu(\theta)q(a\mid\theta)$ is a \textit{Bayes correlated equilibrium} of $E$. I write $\mathcal Q^O(E)$ for the obedient direct devices.

Let $\K(\R)$ denote the nonempty compact subsets of $\R$, equipped
with the Hausdorff distance $d_H$. The principal evaluates a device
$\lambda$ by $\V(S(\lambda))$. I assume throughout that
$\V\colon\K(\R)\to\R$ is continuous and \textit{internal}:
\begin{equation}\label{eq:internality}
 \min S\leq\V(S)\leq\max S
 \qquad(S\in\K(\R)).
\end{equation}
Note that this means $\V(\{s\})=s$.

The principal's optimal values over all devices and over obedient direct devices are\footnote{Both classes are nonempty: agents can receive uninformative signals and play a Nash equilibrium of the game in which I average each utility function over $\mu$. The product distribution of that equilibrium defines an obedient direct experiment independent of the state. Since $v$ is bounded, internality of $\V$ makes both values finite.} \(W(E)\coloneqq\sup_{\lambda}\V(S(\lambda))\) and \(W^O(E)\coloneqq\sup_{q\in\mathcal Q^O(E)}\V(S(q))\), where $\lambda$ ranges over all devices.

I write \(B(E)\coloneqq \max_{q\in\mathcal Q^O(E)} \sum_{\theta,a}\mu(\theta)q(a\mid\theta)v(\theta,a)\) for the largest payoff the principal can obtain from an obedient equilibrium of an obedient direct device.

A \textit{complete-information environment} is the special case in which $\Theta$ is a singleton, in which case I suppress the state and prior, write $E=((A_i,u_i)_i,v)$, identify a device with its distribution $\lambda\in\Delta(T)$, and write $\CE(E)=\mathcal Q^O(E)$ for the ordinary correlated-equilibrium polytope in $\Delta(A)$. Let $\mathsf{N}(E)$ denote the set of principal payoffs generated by Nash equilibria of $(A_i,u_i)_i$.

\section{Incomplete information}\label{sec:bce}

With incomplete information, the obedience principle holds in every environment if and only if the principal evaluates each device by her favorite equilibrium.

\begin{theorem}\label{thm:bce-characterization}
The following are equivalent.
\begin{enumerate}[nosep,noitemsep]
\item $W^O(E)=W(E)$ for every environment.
\item In every environment, for every device $\lambda$
there is an obedient direct device $q$ such that
$\V(S(q))\geq\V(S(\lambda))$.
\item $\V(S)=\max S$ for every nonempty compact $S\subset\R$.
\end{enumerate}
Otherwise, there is an environment with two agents
and a device $\lambda$ with a unique equilibrium such that \(\sup_{q\in\Delta(A)^\Theta}\V(S(q))
 <\V(S(\lambda))=W(E)\).
\end{theorem}
When these conditions hold, both optima are attained and equal the
largest principal payoff $B(E)$ among Bayes correlated equilibria. That (iii) implies the other statements is standard. Likewise, \textup{(ii)} implies \textup{(i)} is immediate because every obedient direct device is indeed a device.

Two devices in the same environment are \textit{relabelings} of one another if, for each agent $i$, there is a
bijection $\pi_i$ between their signal sets such that \(\lambda'(\pi(t)\mid\theta)=\lambda(t\mid\theta)\) for every $(t,\theta)$, where $\pi(t)=(\pi_i(t_i))_i$.
\begin{lemma}\label{lem:bce-construction}
Fix a finite set $F=\{b=a_0>a_1>\cdots>a_k\}$ with $k\geq1$.
There is an environment with two agents, an obedient direct
device $q^0$, and a device $\lambda^+$ with these properties.
\begin{enumerate}[nosep,noitemsep]
\item The principal's payoff at every state and action profile is
at most $b$.
\item $S(q^0)=F$, and for each member of $F$, there is a strict
pure response equilibrium that delivers that payoff.
\item $\lambda^+$ has a unique response equilibrium, which is
strict and delivers payoff $b$.
\item Every direct device with a response equilibrium yielding payoff $b$ is a relabeling of \(q^0\).
\end{enumerate}
Moreover, for every $\eta>0$, there is $\delta>0$ such that every
direct device $q$ satisfies
\begin{equation}\label{eq:bce-near-maximum}
 \max S(q)>b-\delta
 \quad\Longrightarrow\quad d_H(S(q),F)<\eta.
\end{equation}
\end{lemma}

\begin{proof}
Let $N=k+1$ and let both agents' action sets be $\{0,\ldots,k\}$.
The state is $\theta=(t_1,t_2,z)$, where $t_1,t_2$ are independent
and uniform on $\{0,\ldots,k\}$, and $z$ is independent and uniform
on $\{1,2\}$. 

The realized \(t_i\) is agent \(i\)'s \textit{target}. The realized
\(z\) assigns the agents two roles: agent \(z\) receives the first
payoff below, and the other agent receives the second. For agent $i$'s action $x_i$, let $j_i$ be the unique element of
$\{0,\ldots,k\}$ congruent to $x_i-t_i$ modulo $N$. Set
\[
 u_i(\theta,x)=
 \begin{cases}
 -j_i^2,&z=i,\\
 j_i^2-2(j_i-j_{-i})^2,&z\ne i.
 \end{cases}
\]
Define a symmetric matrix $W$ by
\[
 W_{jj}=a_j\quad(0\leq j\leq k),\qquad
 W_{j,j+1}=W_{j+1,j}=\frac{3a_{j+1}-a_j}{2}
 \quad(0\leq j<k),
\]
and set all remaining entries equal to $a_k$. The principal receives
$v(\theta,x)=W_{j_1j_2}$. Only $W_{00}$ equals $b$: every other
diagonal entry is below $b$, and each specified off-diagonal entry
is below the smaller adjacent diagonal entry. Accordingly, the principal never receives more than $b$ and receives $b$ precisely when $x_1=t_1$ and $x_2=t_2$.

Consider the direct device $q^0$ that recommends $x_i=t_i$ to each agent (and does not reveal $z$). For any strategy profile, let \(r_i(j)=\frac1N\sum_{t_i=0}^k \sigma_i((t_i+j)\bmod N\mid t_i)\) be the unconditional distribution of agent $i$'s difference. Conditional on his target, agent $i$ assigns equal probability to the two roles. If he chooses difference $j_i$, the terms $-j_i^2$ and $j_i^2$ cancel when I average his utilities across those roles, leaving expected utility
$-\mathbb E[(j_i-j_{-i})^2]$, where the expectation is over his
opponent's difference. In equilibrium each difference in the
support of $r_i$ must be optimal against $r_{-i}$.

When the opponent's difference has mean $m$, the terms in agent $i$'s expected utility that depend on his own difference $j$ are $-j^2+2jm$, and so his best replies are the integers closest to $m$. If either agent uses a pure difference $j$, the other's unique best reply is $j$. Otherwise each
agent's mixture has exactly two adjacent integers
in its support. When their supports are $\{j,j+1\}$ and
$\{\ell,\ell+1\}$, the first agent's indifference requires the
second agent's mean to be $j+1/2$, which mean lies strictly between
$\ell$ and $\ell+1$, which forces $\ell=j$. Both indifference
conditions then require equal probabilities on $j,j+1$.

If both agents choose the same difference $j$ with probability one, the principal receives $a_j$. If both randomize equally over two adjacent differences, she receives
\[
 \frac14a_j+\frac14a_{j+1}
  +\frac12\frac{3a_{j+1}-a_j}{2}=a_{j+1}.
\]
Hence, every equilibrium payoff belongs to $F$. Conversely, fix any
$j$ and let each agent choose his target plus $j$ modulo $N$ at
every signal. Any deviation to a different difference lowers his
expected utility by at least one. These strict pure equilibria
attain every member of $F$. $j=0$ is obedience.

Under device $\lambda^+$, agent $i$ observes $(t_i,z)$. At each realized role, agent $z$ strictly prefers difference zero regardless of the other agent's behavior. The other agent then faces difference zero and has utility $j_i^2-2j_i^2=-j_i^2$, so he too strictly prefers difference zero. Thus, in the unique response equilibrium, each agent chooses his target at every signal, and the principal receives $b$. The only additional information relative to $q^0$ is the public announcement of the role $z$.

Let $q$ be any direct device and let a response equilibrium give
the principal payoff $b$, in which case it is easy to see that $q$ must be a relabeling of $q^0$. Suppose \eqref{eq:bce-near-maximum} fails. Then, for some
$\eta>0$, there are direct devices $q_m$ with
$\max S(q_m)>b-1/m$ and $d_H(S(q_m),F)\geq\eta$.
Choose $\sigma_m\in\Eq(q_m)$ attaining $\max S(q_m)$. The spaces of direct devices and response strategy profiles are compact. I can consequently pass to a subsequence along which the devices $q_m$ converge to a direct device $q^*$ and their equilibrium response profiles $\sigma_m$ converge to a response strategy profile $\sigma^*$. The inequalities
\eqref{eq:bce-response} are continuous in both arguments, so
$\sigma^*\in\Eq(q^*)$. Continuity of the principal's payoff and
the bound $v\leq b$ imply $\Pi_v(q^*,\sigma^*)=b$.
Moreover, $q^*$ differs from $q^0$ only in the
labels of its signals. Therefore, $S(q^*)=F$, and for each member
of this set, a strict pure equilibrium delivers that payoff.

Each signal under $q^*$ has probability $1/N$. Fix a pure equilibrium in which both agents choose the same difference. Its finitely many strict incentive inequalities continue to hold under $q_m$ for all
sufficiently large $m$. The same strategy profile is, consequently, an
equilibrium under $q_m$, and its payoff converges to the corresponding
member of $F$. Since $F$ is finite, \(\max_{y\in F}\operatorname{dist}(y,S(q_m))\rightarrow 0\).

It remains to show that every equilibrium payoff under $q_m$ lies close to $F$. Otherwise, choose response equilibria whose payoffs remain a strictly positive distance from $F$ and pass to a convergent subsequence of their strategies. The limit strategies satisfy the equilibrium inequalities under $q^*$ because these inequalities are continuous in the device and the strategies. As a consequence, the limit payoff belongs to $F$. Continuity
of the principal's payoff then contradicts the choice of equilibria
whose payoffs stay a strictly positive distance from $F$. Hence, \(\max_{y\in S(q_m)}\operatorname{dist}(y,F)\rightarrow 0\) as well, contradicting $d_H(S(q_m),F)\geq\eta$ and proving \eqref{eq:bce-near-maximum}.
\end{proof}

\subsection{Necessity}
I prove the contrapositive of \textup{(i)}$\implies$\textup{(iii)} and the final statement in \Cref{thm:bce-characterization}. 

Suppose $\V(S)<\max S$ for some nonempty compact set $S$.
Choose finite subsets of $S$ containing $\max S$ and converging
to $S$ in Hausdorff distance. By continuity of $\V$, one such finite set $F$ satisfies $\V(F)<b=\max F$. Because internality of $\V$ implies $\V(\{b\})=b$, $F$ is not a singleton, so I can apply \Cref{lem:bce-construction}.

Set $\gamma=(b-\V(F))/3>0$. By continuity of $\V$ at $F$, there
is $\eta>0$ such that
\[
 d_H(K,F)<\eta\quad\Longrightarrow\quad
 \V(K)<\V(F)+\gamma=b-2\gamma.
\]
Choose $\delta>0$ from \eqref{eq:bce-near-maximum}. Every direct
device $q$ with $\max S(q)>b-\delta$ then has value less than
$b-2\gamma$. Every remaining direct device has value at most
$\max S(q)\leq b-\delta$ by internality of $\V$. Consequently,
\[
 \sup_{\text{direct }q}\V(S(q))
 \leq \max\{b-2\gamma,b-\delta\}<b.
\]
The device $\lambda^+$ has payoff set $\{b\}$ and, therefore,
value $b$. Since $v(\theta,a)\leq b$ at every state and action profile, internality of $\V$ also implies $W(E)\leq b$. Thus,
$W^O(E)<W(E)=b$, contradicting condition~\textup{(i)} and completing the proof of \Cref{thm:bce-characterization}.

\section{Complete Information and Payoff Monotonicity}
\label{sec:ce-characterization}

I now restrict attention to complete-information environments, with no random state. I also impose the additional property for the principal's evaluation that improving her payoff at every action profile cannot lower her evaluation when the agents' utilities and the device stay fixed:
\begin{axiom}[Payoff Monotonicity]\label{ax:ce-payoff-monotonicity}
If two principal payoff functions $v,w\colon A\to\R$ satisfy $v(a)\leq w(a)$ for every $a\in A$, then $\V(S_v(\lambda))\leq\V(S_w(\lambda))$.
\end{axiom}

The agents face the same game under $v$ and $w$, so they have the
same equilibria. Each equilibrium strategy profile delivers the
principal a weakly higher expected payoff under $w$, and the axiom
requires her evaluation to weakly increase as well.

\begin{theorem}\label{thm:ce-monotone-characterization}
Suppose $\V$ satisfies \Cref{ax:ce-payoff-monotonicity}. The following are equivalent:
\begin{enumerate}[nosep,noitemsep,ref=\roman*]
\item\label{item:ce-equality}
$W^O(E)=W(E)$ for every complete-information environment $E$.
\item\label{item:ce-replacement}
In every complete-information environment $E$, for every device $\lambda$ there is a $q\in\CE(E)$ such that
$\V(S(q))\geq\V(S(\lambda))$.
\item\label{item:ce-representation}
There is an increasing function
$h\colon\R\to[-\infty,\infty)$, continuous in the extended-real
topology\footnote{\textit{Viz.}, $b_n\to b$ \(\implies\) $h(b_n)\to h(b)$, where convergence to a finite value has its usual meaning and $h(b_n)\to-\infty$ means that, for every $C\in\R$, $h(b_n)<C$ for all sufficiently large $n$.} and satisfying $h(b)\leq b$ for every $b$, such that
\begin{equation}\label{eq:ce-characterized-evaluation}
 \V(S)=\max\{\min S,h(\max S)\}
 \qquad\text{for every }S\in\K(\R).
\end{equation}
\end{enumerate}
The function $h$ is unique and satisfies
$h(b)=\inf_{a\leq b}\V(\{a,b\})$.
\end{theorem}

Conversely, every function $h$ with the stated properties defines, through \eqref{eq:ce-characterized-evaluation}, a continuous internal evaluation satisfying payoff monotonicity and the equivalent conditions.

\begin{corollary}\label{cor:ce-optimal-value}
If the equivalent conditions in \Cref{thm:ce-monotone-characterization} hold, \(W(E)=W^O(E)=\max\{\min \mathsf{N}(E),h(B(E))\}\).
\end{corollary}

\begin{proposition}\label{prop:ce-strict-failure}
If the equivalent conditions in
\Cref{thm:ce-monotone-characterization} fail, there is a complete-information environment with three agents, two of whom have
binary action sets, and a device $\lambda$ with $|S(\lambda)|=2$ such that \(\sup_{q\in\Delta(A)}\V(S(q)) <\V(S(\lambda))=W(E)\).
\end{proposition}

Examples of evaluations of the form specified in \eqref{eq:ce-characterized-evaluation} include the minimum, obtained by taking $h\equiv-\infty$, and the maximum, obtained by taking $h(b)=b$. They also include evaluations such as $\max\{\min S,\max S-c\}$ for $c\geq0$, and $\max\{\min S,\min\{\max S,k\}\}$ for a fixed $k\in\R$.

Here are three properties of the complete-information setting that explain why a pessimistic principal loses nothing by moving to obedient devices.
\begin{lemma}\label{lem:ce-equilibrium-facts}
Fix an environment $E$. Then,
\begin{enumerate}[nosep,noitemsep]
\item Every device $\lambda$ satisfies $\mathsf{N}(E)\subseteq S(\lambda)$.
\item The action distribution induced by any response equilibrium of any device belongs to $\CE(E)$.
\item If $p=\prod_i p_i$ is the action distribution of a Nash
      equilibrium, then the direct device $p$ is obedient and
      $S(p)=\mathsf{N}(E)$.
\end{enumerate}
\end{lemma}

\begin{proof}
(i) follows because each agent can simply disregard the device and assume all others are, and (ii) and (iii) are similarly immediate.\end{proof}

$\mathsf{N}(E)$ is nonempty and compact, and $\CE(E)$ is a nonempty compact polytope, so a correlated equilibrium attains $B(E)$. These facts and \Cref{lem:ce-equilibrium-facts} imply, for every device $\lambda$, \(\min S(\lambda)\leq\min \mathsf{N}(E)\) and \(\max S(\lambda)\leq B(E)\).

\begin{lemma}\label{lem:ce-endpoints}
The evaluation $\V$ satisfies payoff monotonicity if and only if
there is a continuous function $f$ on
$\{(m,M)\in\R^2\colon m\leq M\}$, increasing in each argument and
satisfying $m\leq f(m,M)\leq M$, such that \(\V(S)=f(\min S,\max S)\) for every \(S\in\K(\R)\). In that case, $f(m,M)=\V(\{m,M\})$.
\end{lemma}

\begin{proof}
Suppose first that $\V$ satisfies payoff monotonicity. Fix $k \in \mathbf N$, give each of two agents actions $1,\ldots,k$, and set \(u_1(i,j)=\mathbf 1\{i=k\text{ or }i=j\}\) and \(u_2(i,j)=\mathbf 1\{j=k\text{ or }i=j\}\).

The Nash equilibria are precisely the pure profiles $(i,i)$.\footnote{Indeed, action $k$ guarantees payoff one. If agent 1 uses $i<k$ with strictly positive probability, his optimality requires agent 2 to choose
$i$. Agent 2's optimality then requires agent 1 to choose
$i$. If agent 1 only uses $k$, agent 2's unique best response
is $k$. Each diagonal profile is itself an equilibrium.} Use a device that gives each agent a constant signal. Given vectors $x,y\in\R^k$ with $x_i\leq y_i$ for every $i$, I specify two payoff functions for the principal $v$ and $w$ by $v(i,i)=x_i$, $w(i,i)=y_i$, and $v(i,j)=w(i,j)=0$ whenever $i\neq j$. Payoff monotonicity of \(\V\) implies
\begin{equation}\label{eq:ce-coordinate-monotonicity}
 \V(\{x_1,\ldots,x_k\})\leq\V(\{y_1,\ldots,y_k\}).
\end{equation}

Now take any finite set $\{x_1,\ldots,x_k\}$ with
$m=x_1\leq\cdots\leq x_k=M$ and $k\geq2$. \eqref{eq:ce-coordinate-monotonicity} and \((m,\ldots,m,M)\leq(x_1,\ldots,x_k)\leq(m,M,\ldots,M)\) imply $\V(\{x_1,\ldots,x_k\})=\V(\{m,M\})$. If the set is the singleton $\{m\}$, then $M=m$ and $\{m,M\}=\{m\}$, and so internality of $\V$ implies $\V(\{m\})=m$. Define $f(m,M)\coloneqq\V(\{m,M\})$. Apply \eqref{eq:ce-coordinate-monotonicity} to get that $f$ increases in each argument, and continuity and internality of $\V$ imply continuity and internality of $f$. For any nonempty compact $S$, choose finite sets $S_n$ containing $\min S$ and $\max S$ and converging to $S$ in the Hausdorff distance. We have shown that $\V(S_n)=\V(\{\min S,\max S\})$ for every $n$, so continuity of $\V$ delivers $\V(S)=f(\min S,\max S)$.

Conversely, suppose \(\V(S)=f(\min S,\max S)\) for every \(S\in\K(\R)\) holds for such
an $f$. Fix the agents' utilities and a device $\lambda$, and
increase the principal's payoff from $v$ to $w$ pointwise. The
equilibria in the game do not change, and
$\Pi_v(\lambda,\sigma)\leq\Pi_w(\lambda,\sigma)$ for each
$\sigma\in\Eq(\lambda)$, implying that both the minimum and the maximum over this set weakly increase. Given this, monotonicity of $f$ implies payoff monotonicity of $\V$.
\end{proof}

\begin{lemma}\label{ce:construction}
Fix $c<a<b$. There is an environment $E$ with three agents, two of whom have binary action sets, and a device $\lambda$ such that the following hold:
\begin{enumerate}[nosep,noitemsep]
\item $\mathsf{N}(E)=\{a\}$ and $S(\lambda)=\{a,b\}$.
\item Every principal payoff generated by a response equilibrium under any device belongs to $[c,b]$.
\item Some obedient direct device $q\in\CE(E)$ satisfies
$\max S(q)=b$.
\item For every $\eta>0$, there is $\delta>0$ such that every direct
device $q\in\Delta(A)$ satisfies
\(\max S(q)>b-\delta\) \(\implies\) \(\min S(q)<c+\eta\).
\end{enumerate}
\end{lemma}
\begin{proof}

I first specify the game played by agents 1 and 2. Each has action set \(\{x,y\}\), and their payoffs are
\[
\begin{array}{c|cc}
 & x & y\\ \hline
x & (4/3,4/3) & (7,2)\\
y & (2,7) & (5,5).
\end{array}
\]
This is a standard anti-coordination game: each agent strictly prefers $x$ when the other agent's probability of $x$ is below $3/4$, strictly prefers $y$ when that probability exceeds $3/4$, and is indifferent at $3/4$. The three Nash equilibria are \((x,y)\), \((y,x)\), and the symmetric mixed equilibrium in which each agent chooses \(x\) with probability \(3/4\).

Consider the direct recommendation distribution $q^*$, whose rows detail agent 1's recommendation and whose columns detail agent 2's, both ordered $x,y$:
\[
 q^*=\begin{pmatrix}1/8&7/16\\7/16&0\end{pmatrix}.
\]
There is a strict response equilibrium in which each agent follows his recommendation. There is also a strict response equilibrium in which each agent chooses $y$ after recommendation $x$ and chooses $x$ after recommendation $y$. Conditional on an agent receiving recommendation $x$, the other agent then chooses $x$ with probability $7/9$; conditional on the agent receiving recommendation $y$, the other agent chooses $x$ with probability zero. This response equilibrium induces the action distribution $q^-$:
\[
 q^-=\begin{pmatrix}0&7/16\\7/16&1/8\end{pmatrix}.
\]

Now consider the joint signal distribution $\lambda_0$:\footnote{The game played by agents 1 and 2 and the family containing their signal device $\lambda_0$ are \citet[Example~2.1 and Figure~4]{Ray2002}. I will add a third agent below in order to generate the required set of equilibrium payoffs for the principal.}
\[
 \lambda_0=
 \begin{pmatrix}
 0&0&1/7\\
 1/8&7/16&0\\
 33/112&0&0
 \end{pmatrix}.
\]
Each agent's signal identifies one of two components of this
distribution. The isolated signal pair, with probability $1/7$,
induces the original two-agent game. The remaining component has
unnormalized distribution
\[
 M=\frac1{112}\begin{pmatrix}14&49\\33&0\end{pmatrix}.
\]
Let $x_1,x_2$ denote agent 1's probabilities of choosing $x$ following
the two signals in this component, and let $y_1,y_2$ denote agent 2's
probabilities. Write $\beta(p)$ for the optimal probabilities of choosing $x$ against
an agent who chooses $x$ with probability $p$; thus $\beta(p)=\{1\}$
for $p<3/4$, $\beta(3/4)=[0,1]$, and $\beta(p)=\{0\}$ for $p>3/4$.
At equilibrium, it must be that
\[
 x_2\in\beta(y_1),\quad y_2\in\beta(x_1),\quad
 x_1\in\beta\left(\frac{2y_1+7y_2}{9}\right),\quad \text{\&} \quad
 y_1\in\beta\left(\frac{14x_1+33x_2}{47}\right).
\]
The response profiles satisfying these four equilibrium conditions
appear in the table below. The last column states this component's contribution to the unconditional probability of $(y,y)$, which, for a response profile $(x_1,x_2,y_1,y_2)$, is
\[
 \frac{14(1-x_1)(1-y_1)+49(1-x_1)(1-y_2)+33(1-x_2)(1-y_1)}{112}.
\]
And here are the response profiles satisfying the four equilibrium conditions:
\[
\begin{array}{cccc|c}
 x_1&x_2&y_1&y_2&\text{contribution to }\Prb(y,y)\\\hline
 0&0&1&1&0\\
 1&1&0&0&0\\
 1&0&1&0&0\\
 1&85/132&3/4&0&47/1792\\
 3/4&1&0&27/28&63/1792\\
 3/4&0&1&19/28&63/1792\\
 3/4&3/4&3/4&3/4&3/56
\end{array}
\]

Let $\mathcal R$ be the set of action distributions induced by response equilibria under $\lambda_0$. Every $Q\in\mathcal R$ satisfies
\[
 Q(y,y)\le\frac{3}{56}+\frac1{7}\frac1{16}=\frac1{16}.
\]
The same array $q^*$, now viewed as a distribution over action
profiles, belongs to $\mathcal R$. Use the third response profile in
the table in the $M$ component and play $(y,x)$ at the isolated signal
pair. Listing action profiles in the order
$(x,x),(x,y),(y,x),(y,y)$, the induced distribution is
\[
 \frac1{112}(14,49,33,0)+\frac17\delta_{(y,x)}
 =\frac1{112}(14,49,49,0)=q^*,
\]
where $\delta_z$ denotes the degenerate distribution on action profile
$z$. The agents can also ignore their signals and play any of the three
Nash equilibria, so \(\delta_{(x,y)},\delta_{(y,x)},\nu\in\mathcal R\) where $\nu=(9,3,3,1)/16$ is the action distribution in the mixed Nash
equilibrium.

Write $\Delta_{12}\coloneqq
\Delta(\{(x,x),(x,y),(y,x),(y,y)\})$. Because $\mathcal R$ is finite, I can choose $\varepsilon\in(0,1/32)$ so that the closed ball in the sup norm of radius
$\varepsilon$ around $q^*$ contains no other member of $\mathcal R$. For
$P\in\Delta_{12}$, define
\[d_+(P)=\frac{\|P-q^*\|_\infty}{\varepsilon},\quad
 d_-(P)=\frac{\|P-q^-\|_\infty}{\varepsilon}, \quad \text{\&} \quad
 t_\pm(P)=\max\{1,d_\pm(P)\}-d_\pm(P),
\]
and set \(F(P)=a+(b-a)t_+(P)-(a-c)t_-(P)\).

The supports of the two functions $t_+,t_-$ are disjoint because
$\|q^*-q^-\|_\infty=1/8$. Also, $t_-$ vanishes throughout $\mathcal R$: every member of $\mathcal R$ assigns \((y,y)\) probability at most $1/16$, whereas $q^-(y,y)=1/8$. Consequently, \(F(q^-) = c\leq F(P)\leq b\) and \(F(P)=b\ \Longleftrightarrow\ P=q^*\). Since $t_+$ also vanishes on
$\mathcal R\setminus\{q^*\}$, $F(Q)=a$ for every
$Q\in\mathcal R\setminus\{q^*\}$. The three Nash-equilibrium
distributions $\delta_{(x,y)},\delta_{(y,x)},\nu$ belong to this set, so $F$ equals $a$ at each of them.

I now add a third agent so that, whenever the first two agents' actions
have distribution $P$, each of his best responses gives the principal
payoff $F(P)$. To construct his payoffs, define
$p_\pm(P)=\max\{1,d_\pm(P)\}$ and write $F=g_1-g_2$, where
\[
 g_1=a+(b-a)p_++(a-c)d_-,\qquad \text{\&} \qquad
 g_2=(b-a)d_++(a-c)p_-.
\]
Note that I can write
$g_1(P)=\max_{r\in I}\ell_r(P)$ and
$g_2(P)=\max_{s\in J}m_s(P)$, for finite sets $I,J$ and affine
functions $\ell_r,m_s$.

Give agent 3 action set $I\times J$ and define, at action profile
$(z,(r,s))$, his utility and the principal's payoff by
\[
 u_3(z,(r,s))=\ell_r(\delta_z)+m_s(\delta_z),\qquad \text{\&} \qquad
 v(z,(r,s))=\ell_r(\delta_z)-m_s(\delta_z).
\]
Agents 1 and 2 retain their original utilities. Against distribution $P$, agent 3 maximizes
$\ell_r(P)+m_s(P)$. Because his action set is $I\times J$, an action
$(r,s)$ is optimal for him exactly when $r$ maximizes $\ell_r(P)$ and
$s$ maximizes $m_s(P)$. Every such action, and every mixture
of such actions, gives the principal expected payoff
$g_1(P)-g_2(P)=F(P)$.

I give agents 1 and 2 device $\lambda_0$ and agent 3 a constant
signal. The first two agents' equilibrium action distributions range over $\mathcal R$, and agent 3's best responses deliver $F(\mathcal R)=\{a,b\}$. In every Nash equilibrium of the resulting three-agent game, the first
two agents play one of the three Nash equilibria of their two-agent
game, so agent 3's best response gives the principal payoff $a$.

Now fix any device and any response equilibrium. For each signal of
agent 3 that occurs with strictly positive probability, let $P$ be the
conditional distribution of the first two agents' actions given that
signal. Because agent 3 best responds after that signal, the
principal's expected payoff conditional on the signal is
$F(P)\in[c,b]$. The principal's \textit{ex ante} payoff is the average of these conditional payoffs over agent 3's signals, so it also belongs to $[c,b]$.
Finally, recommend $q^*$ to agents 1 and 2 and give agent 3 a constant
recommendation that maximizes his utility against $q^*$. This direct
device is obedient and gives the principal payoff $b$ under obedience.

Suppose a response equilibrium of a direct device $q$ gives the
principal payoff $b$. Conditional on each signal of agent 3 that occurs with strictly positive probability, let $P$ be the conditional distribution of the first two agents' actions. The principal's conditional payoff is $F(P)\leq b$. Because these conditional payoffs average to $b$ and $q^*$ uniquely maximizes $F$, each such conditional distribution is $q^*$. Their unconditional action distribution is therefore also $q^*$.

Neither of the first two agents randomizes following a signal that
occurs with strictly positive probability and, since each chooses both $x$ and $y$ with strictly positive probability under $q^*$, each chooses $x$ after one of his two signals and $y$ after the other. For each of these agents, relabel the signal that induces $x$ as $x$ and the signal that induces $y$ as $y$. Conditional on each signal of agent 3 that occurs with strictly positive probability, the distribution of the first two agents' signals is then $q^*$. Hence $q(t_1,t_2,t_3)=q^*(t_1,t_2)p_3(t_3)$, where $p_3$ is the marginal distribution of agent 3's signal. If both agents interchange $x$ and $y$, they play a strict response equilibrium and induce $q^-$, independently of agent 3's signal. Let agent 3 choose a best response at each signal. Together, these strategies form a response equilibrium with principal payoff $F(q^-)=c$.

For the final assertion, suppose there were $\eta>0$ and direct devices $q_j$ with $\max S(q_j)\to b$ and $\min S(q_j)\ge c+\eta$.
Choose $\sigma_j\in\Eq(q_j)$ attaining each maximum. After passing to a subsequence, take $q_j\to q_\infty$ and $\sigma_j\to\sigma_\infty$. The expressions on the left of \eqref{eq:bce-response}
are continuous in the device and strategy profile, so
$\sigma_\infty\in\Eq(q_\infty)$. Continuity of the principal's
expected payoff implies that this equilibrium attains $b$.

Relabel the first two agents' signals according to their actions under $\sigma_\infty$, as above, and apply the same relabeling to $q_\infty$ and every $q_j$. This does not change any payoff set. Then $q_\infty(t_1,t_2,t_3)=q^*(t_1,t_2)p_3(t_3)$. Under $q_\infty$, both labels of each of the first two agents occur with strictly positive probability. The strict inequalities in incentives for the response that interchanges $x$ and $y$ thus continue to hold under $q_j$ for all sufficiently large $j$. For each such $j$, let the first two agents use this response and let agent 3 choose a best response at each of his signals. These strategies form a response equilibrium of $q_j$ because the first two agents' utilities do not depend on agent 3.

At each signal $t_3$ of agent 3 with $p_3(t_3)>0$, the conditional distribution of the first two agents' actions given $t_3$ converges to $q^-$. Continuity of $F$ implies that the principal's expected payoff conditional on $t_3$ converges to $c$. The signals $t_3$ with $p_3(t_3)=0$ have total probability converging to zero, and their conditional principal payoffs remain in $[c,b]$. The \textit{ex ante} principal payoff of these response equilibria thus converges to $c$, contradicting $\min S(q_j)\ge c+\eta$.
\end{proof}

\subsection{Completing the proof of necessity}

\begin{proof}[Proof of \textup{(\ref{item:ce-equality})}$\implies$\textup{(\ref{item:ce-representation})} in \Cref{thm:ce-monotone-characterization}]

Suppose the two suprema coincide in every environment. By
\Cref{lem:ce-endpoints}, I can write
$\V(S)=f(\min S,\max S)$, where $f$ is continuous, internal, and
increasing in each argument. I first show that
\begin{equation}\label{ce:lower-endpoint}
 c<a<B,\quad f(a,B)>a
 \quad\Longrightarrow\quad f(c,B)=f(a,B).
\end{equation}
Since $f$ increases in its first argument, it suffices to rule out
$f(c,B)<f(a,B)$.

Suppose this strict inequality holds, and write $r=f(a,B)>a$.
Let $b$ be the smallest element of $[a,B]$ such that $f(a,b)=r$.
Continuity of $f$ and $f(a,a)=a<r=f(a,B)$ guarantee that $b$ exists
and belongs to $(a,B]$. Because $f$ increases in its second argument,
$f(a,z)<r$ for every $z\in[a,b)$.

Apply \Cref{ce:construction} to $c<a<b$. In the resulting environment, every device $\lambda'$ satisfies
$S(\lambda')\subseteq[c,b]$ and $a\in S(\lambda')$, while the device
in the lemma satisfies $S(\lambda)=\{a,b\}$. The principal consequently
assigns this device value $r$. Moreover,
monotonicity of $f$ bounds the value of every device by
$f(a,b)=r$. Hence $W(E)=r$.

I now bound the value of every direct device strictly below
$r$. Since $b\le B$, monotonicity of $f$ implies
$f(c,b)\le f(c,B)<r$. By continuity of $f$, I can choose
$\eta\in(0,a-c)$ such that $f(c+\eta,b)<r$.
\Cref{ce:construction} supplies $\delta\in(0,b-a)$ such that
every direct device $q$ satisfies
\[
 \max S(q)>b-\delta
 \quad\Longrightarrow\quad
 \min S(q)<c+\eta.
\]
For such a device, monotonicity of $f$ implies
$\V(S(q))\le f(c+\eta,b)<r$. Every remaining direct device has
$\min S(q)\le a$ and $\max S(q)\le b-\delta$, so
$\V(S(q))\le f(a,b-\delta)<r$. Thus,
\[
 \sup_{q\in\Delta(A)}\V(S(q))
 \le \max\{f(c+\eta,b),f(a,b-\delta)\}<r,
\]
contradicting equality of the suprema and proving
\eqref{ce:lower-endpoint}.

For each $B\in\R$, define $h(B)=\inf_{m\le B}f(m,B)$.
For every $m\leq B$, internality of $\V$ implies $f(m,B)=\V(\{m,B\})\leq B$, so $h(B)\leq B$. If $f(m,B)=m$, then $h(B)\le m$,
so $f(m,B)=\max\{m,h(B)\}$. If $f(m,B)>m$, \eqref{ce:lower-endpoint} implies $f(c,B)=f(m,B)$ for every $c<m$.
Since $f$ increases in its first argument, its infimum over all
$c\le B$ is $f(m,B)$. This again yields \(f(m,B)=\max\{m,h(B)\}\). The representation also identifies $h$ uniquely, since \(\inf_{m\le B}\max\{m,h(B)\}=h(B)\).

For $B<B'$, monotonicity of $f$ implies $f(m,B)\le f(m,B')$ whenever
$m\le B$. Letting $m\to-\infty$ shows that $h(B)\le h(B')$.
To check continuity at a point $B_0$ with $h(B_0)>-\infty$, choose
$m<h(B_0)$. Then $f(m,B_0)=h(B_0)>m$. By continuity of $f$,
$f(m,B)>m$ for $B$ near $B_0$, so the representation implies
$h(B)=f(m,B)$ there. Thus, $h$ is continuous at $B_0$.
If $h(B_0)=-\infty$, take any $C\in\R$ and choose
$m<\min\{C,B_0\}$. Since $f(m,B_0)=m<C$, continuity of $f$ implies
$h(B)\le f(m,B)<C$ for $B$ near $B_0$. This proves continuity in
the extended-real topology.

If the representation in the theorem fails, then
\eqref{ce:lower-endpoint} must fail, since the preceding argument
obtains the representation from that identity. The same construction
then proves \Cref{prop:ce-strict-failure}.
\end{proof}

\subsection{Sufficiency and the optimal value}

\begin{proof}[Proof of \textup{(\ref{item:ce-representation})}\(\implies\)\textup{(\ref{item:ce-replacement})}\(\implies\)\textup{(\ref{item:ce-equality})} in \Cref{thm:ce-monotone-characterization}]

Conversely, take a function $h$ with the stated properties and define
$\V$ by \eqref{eq:ce-characterized-evaluation}.
This evaluation is internal and increasing in both endpoints.
Since $h$ is continuous, $\V$ is Hausdorff continuous at any set
for which $h(\max S)$ is finite. If
$h(\max S)=-\infty$, choose a number strictly below $\min S$.
For every compact set sufficiently close to $S$ in Hausdorff distance,
continuity of $h$ in the extended-real topology ensures that its value
at the maximum remains below that number, while the minimum remains
above it. Accordingly, on this neighborhood, $\V$ equals the minimum, which is Hausdorff continuous. Because a pointwise increase in the
principal's payoff increases both endpoints of her payoff set over the
fixed response equilibria, $\V$ satisfies payoff monotonicity.

Fix an environment $E$, write $N=\mathsf{N}(E)$, and set
$a=\min N$ and $B=B(E)$. By \Cref{lem:ce-equilibrium-facts},
every device allows the agents to play each Nash equilibrium, and the action distribution under any response equilibrium is a correlated equilibrium. Thus, the principal's worst equilibrium payoff cannot exceed
$a$, and her best cannot exceed $B$. Since $h$ is increasing,
\(\V(S(\lambda))\le\max\{a,h(B)\}\) for every device $\lambda$.

If $h(B)\le a$, independent recommendations generated by a Nash
equilibrium have payoff set exactly $N$. Since
$h(\max N)\le h(B)\le a$, this device has value $a$.
If $h(B)>a$, choose a correlated equilibrium $q$ attaining $B$.
Obedience delivers $B$, every response payoff is at most $B$, and
$S(q)$ contains $N$. Thus $\max S(q)=B$ and $\min S(q)\le a<h(B)$,
so this direct device has value $h(B)$. In either case an obedient
direct device attains the upper bound and weakly dominates every
device. Both optima thus exist and equal
$\max\{\min N,h(B)\}$.
This proves \textup{(\ref{item:ce-representation})}
\(\implies\)\textup{(\ref{item:ce-replacement})}. Condition
\textup{(\ref{item:ce-replacement})} implies
\textup{(\ref{item:ce-equality})} because obedient direct devices
are indeed devices. Along with the proof of necessity,
this establishes the equivalence of
\textup{(\ref{item:ce-equality})},
\textup{(\ref{item:ce-replacement})}, and
\textup{(\ref{item:ce-representation})}. This also proves \Cref{cor:ce-optimal-value}.
\end{proof}

\bibliography{correq}

\end{document}